\documentclass[iop]{emulateapj}
\usepackage{apjfonts}

\newcommand{\lp}{\left(}
\newcommand{\rp}{\right)}
\newcommand{\E}[1]{\times10^{#1}}
\newcommand{\smpy}{ \ M_\odot \ {\rm yr}^{-1}}
\newcommand{\msol}{ \ M_\odot }
\newcommand{\sne}{SNe\ }
\newcommand{\ia}{Ia\ }

\shorttitle{CSM ABSORPTION IN DOUBLE DETONATION \sne \ia}
\shortauthors{SHEN, GUILLOCHON, \& FOLEY}
\slugcomment{Accepted for publication in The Astrophysical Journal Letters}

\begin{document}

% -----------------------------------------------------------
% -----------------------------------------------------------

\title{Circumstellar absorption in double detonation Type \ia supernovae}

\author{Ken J. Shen\altaffilmark{1,2,5}, James Guillochon\altaffilmark{3}, and Ryan J. Foley\altaffilmark{4,6}}
\altaffiltext{1}{Lawrence Berkeley National Laboratory, 1 Cyclotron Road, Berkeley, CA 94720, USA; kenshen@astro.berkeley.edu}
\altaffiltext{2}{Department of Astronomy and Theoretical Astrophysics Center, University of California, Berkeley, CA 94720, USA}
\altaffiltext{3}{Department of Astronomy and Astrophysics, University of California, Santa Cruz, CA 95064, USA}
\altaffiltext{4}{Harvard-Smithsonian Center for Astrophysics, 60 Garden Street, Cambridge, MA 02138, USA}
\altaffiltext{5}{Einstein Fellow.}
\altaffiltext{6}{Clay Fellow.}

% -----------------------------------------------------------
% -----------------------------------------------------------

\begin{abstract}

Upon formation, degenerate He core white dwarfs are surrounded by a radiative H-rich layer primarily supported by ideal gas pressure.  In this Letter, we examine the effect of this H-rich layer on mass transfer in He+C/O double white dwarf binaries that will eventually merge and possibly yield a Type Ia supernova (SN Ia) in the double detonation scenario.  Because its thermal profile and equation of state differ from the underlying He core, the H-rich layer is transferred stably onto the C/O white dwarf prior to the He core's tidal disruption.  We find that this material is ejected from the binary system and sweeps up the surrounding interstellar medium hundreds to thousands of years before the SN Ia.  The close match between the resulting circumstellar medium profiles and values inferred from recent observations of circumstellar absorption in SNe Ia gives further credence to the resurgent double detonation scenario.

\end{abstract}


% -----------------------------------------------------------
% -----------------------------------------------------------

\section{Introduction}

Type Ia supernovae (SNe Ia) are responsible for nearly half of the heavy element production in the Universe \citep{tww95} and serve as probes of cosmic acceleration \citep{ries98,perl99}.  However, the nature of their progenitor systems is still a mystery.  While researchers agree that SNe Ia involve the explosions of C/O white dwarfs (WDs), three main possibilities for the companion to the exploding WD remain in contention: a H-rich donor (typically referred to as a ``single degenerate'' system; \citealt{wi73,nomo82a}), another C/O WD (``double degenerate''; \citealt{it84,webb84}), and a He WD or He-burning star (``double detonation''; \citealt{livn90}).  Despite concerted effort, no consensus has yet been reached.

Recent theoretical work \citep{schw12,shen12}, in an update of initial studies (e.g., \citealt{ni85}), suggests that the long term evolution of a double degenerate system does not yield a SN Ia but instead leads to the formation of a C-rich giant and then possibly to a collapse to a neutron star \citep{sn85}.  Furthermore, a wide variety of work, both theoretical (e.g., \citealt{sb07,rbf09,kase10}) and observational (e.g., \citealt{dist10a,gb10}), suggests that single degenerate systems cannot be the dominant progenitor channel.  While the double detonation scenario has been comparatively less well-studied, recent work \citep{fhr07,fink10,sim10} has resurrected interest in them and spurred ongoing research.

One point seemingly in favor of single degenerate progenitors has come from recent observations of circumstellar material (CSM) surrounding $10-30\%$ of SNe Ia at distances of $0.1-1$ pc from the explosion and velocities of $50-150$ km s$^{-1}$ \citep{pata07,blon09,simo09,ster11,fole12a}.  The authors have concluded that the presence of CSM this close to the explosion implies some form of a H-rich single degenerate progenitor, since several variants of this scenario predict mass loss histories that might mimic the observed CSM distributions.

However, we show in this Letter that H-rich material is also ejected hundreds to thousands of years prior to the merger of a He WD and a C/O WD, which may lead to a SN Ia in the double detonation scenario.  For ambient interstellar medium (ISM) densities appropriate for spiral galaxies, this ejecta and the swept-up ISM yield a CSM profile that matches observed distances, velocities, and column densities.  We also find that the lower ISM density in elliptical galaxies inhibits the formation of neutral Na in the CSM, which could explain the non-detection of neutral Na absorption in SNe Ia in ellipticals without dust lanes.

We begin in Section \ref{sec:progevol} by describing the binary evolution that produces a He WD with a thin surface H layer.  We derive the properties of the initially stable H-rich mass transfer from the He WD to the C/O WD and perform a time-dependent calculation of this accretion onto the C/O WD in Section \ref{sec:mdot}.  Prior to the merger of the He WD with the C/O WD, the H-rich accretion yields multiple ejection events akin to classical novae, which expel material into the surrounding ISM.  We model the resulting CSM in Section \ref{sec:ISM}, and conclude in Section \ref{sec:conc}.

% -----------------------------------------------------------
% -----------------------------------------------------------

\section{Evolutionary scenario and progenitor characteristics}
\label{sec:progevol}

Double detonations may arise in systems with a C/O WD primary via stable He mass transfer from a He-burning star \citep{livn90} or low mass He WD donor \citep{fhr07}, or by unstable mass transfer from a higher mass He WD \citep{guil10} or C/O WD donor \citep{pkt13a}.  The subclass of double detonation progenitors we consider in this work are binaries with a relatively high mass $0.3-0.45 \msol$ He WD and a $0.9-1.2 \msol$ C/O WD.  This range of He WD masses is predicted to yield a He detonation either via an accretion stream instability in the lead-up to the merger or during the disruption of the He WD \citep{guil10,dan12}, or possibly during the viscous evolution of the merger remnant \citep{schw12}.  Detonations of either lower or higher mass C/O WDs underproduce or overproduce $^{56}$Ni, respectively, as compared to typical SNe Ia \citep{sim10,ruit13a}.  Throughout this Letter, we focus on a fiducial binary with a $0.40 \msol$ He WD and a $1.0 \msol$ C/O WD.  Binaries with different components will yield quantitatively, but likely not qualitatively, different results, which we defer to future work.

We construct our fiducial $0.40 \msol$ He WD with the stellar evolution code MESA\footnote{http://mesa.sourceforge.net/ (version 4589; default settings are used unless otherwise noted)} \citep{paxt11} by truncating the evolution of a $1.0 \msol$ star with an initial metallicity of $Z=0.01$ as it ascends the red giant branch (RGB).  Chemical diffusion is active throughout the calculation.  When the H-deficient core mass reaches $0.399 \msol$, mass loss is turned on to rapidly remove the convective H envelope, yielding a $0.40 \msol$ He core WD.

Upon formation, He WDs possess a thin, $2\E{-4}- 2\E{-3} \msol$ H-rich remnant surface layer.  This layer evolves under the action of cooling, chemical diffusion, and nuclear burning, yielding a radiative layer containing essentially pure H on top of a degenerate He core polluted with $^{14}$N.  We refer readers to previous studies (e.g., \citealt{asb01a,kbs12}) for a comprehensive description of He WD evolution and instead describe our two fiducial He WD models, which evolve for $10^8$ and $3\E{9}$ yr after the truncation of the RGB prior to the onset of mass transfer to the C/O WD.  These two timescales are chosen to parameterize our uncertainties in the double WD binary's separation following common envelope evolution.  The elapsed time between the formation of the He WD and the onset of mass transfer depends strongly on this separation and can range from a Hubble time for an initial separation of $a=1.3\E{11}$ cm to $10^8$ yr for a separation of $3.8\E{10}$ cm.

\begin{figure}
	\plotone{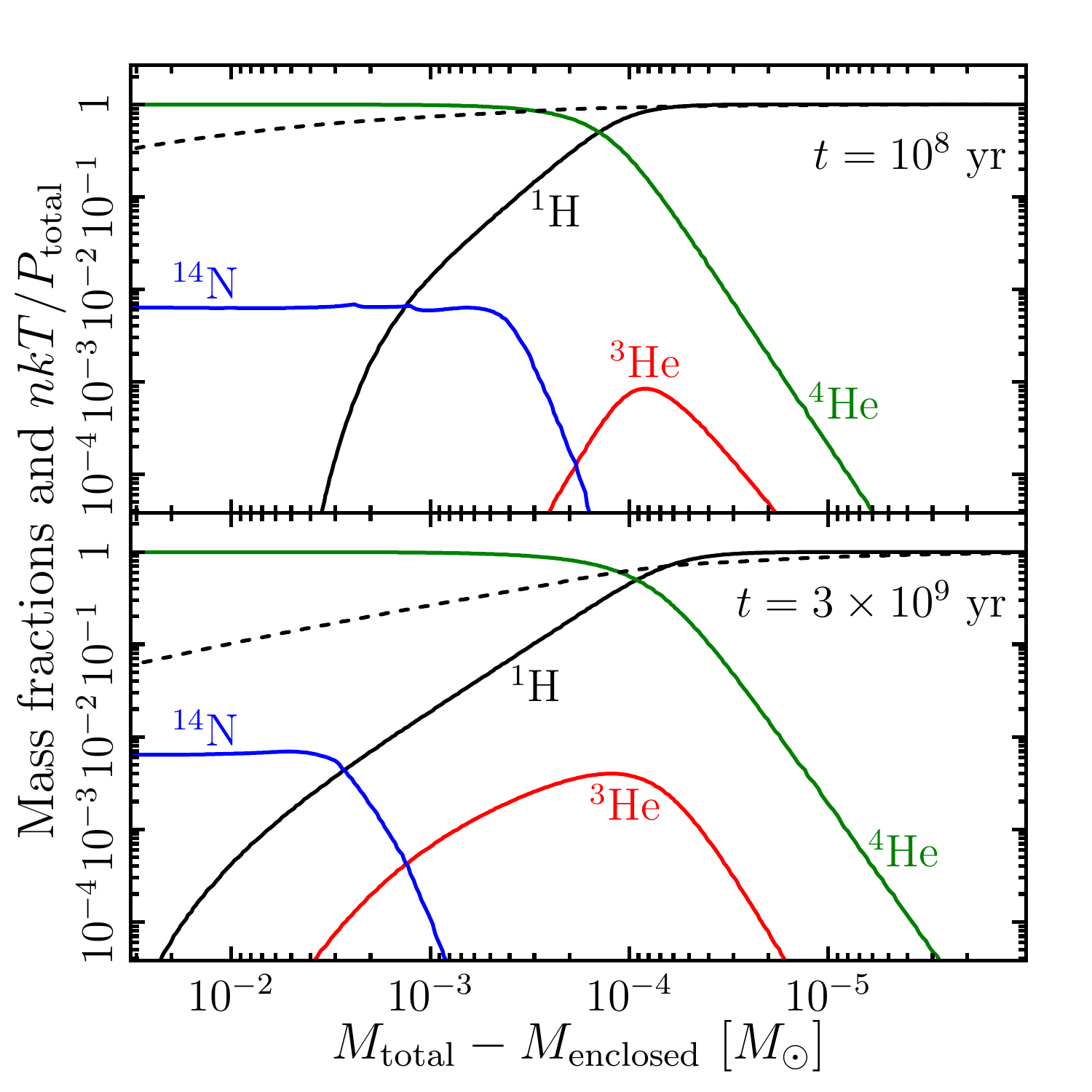}
	\caption{Mass fractions vs.\ mass below surface of a $0.40 \msol$ He WD, $10^8$ yr (\emph{top panel}) and $3\E{9}$ yr (\emph{bottom panel}) after truncating RGB evolution.  Solid lines correspond to mass fractions of $^1$H, $^3$He, $^4$He, and $^{14}$N as labeled.  Also shown as a dashed line in both panels is the ratio of ideal gas pressure, $nkT$, to total pressure, $P_{\rm total}$, where $n$ is the combined number density of ions and free electrons, and $T$ is the temperature.}
	\label{fig:0_4profiles}
\end{figure}

The initial separation and resulting merger timescale are critical for determining the mass transfer history once accretion begins.  The H layer on an older He WD will be colder and thinner, and such a system will have a shorter delay between the onset of H-rich mass transfer and the disruption of the He WD.  The cooling and contraction of the WD, as well as a small amount of residual H-burning, also affect the abundance structure within the surface layers.  Figure \ref{fig:0_4profiles} shows abundance profiles for our two He WDs.  Since $^3$He and $^{14}$N act as catalysts for H-burning in classical and recurrent novae \citep{tb04,sb09a}, a proper calculation of the mass transfer history onto the C/O WD, which we describe in the next section, should take into account both the changing thermal properties and abundance profiles of the aging He WD.

% -----------------------------------------------------------
% -----------------------------------------------------------

\section{Time dependent mass transfer rate and mass ejection history}
\label{sec:mdot}

When mass transfer begins, the He WD's radiative H layer is transferred first.  We follow the He WD donor's response to this mass transfer by removing mass at a constant rate of $10^{-7} \smpy$ in MESA (see \citealt{kbs12} for a similar analysis of lower mass He WDs).  The choice of mass removal rate in this calculation is essentially arbitrary because the thermal timescale in the H layer is $ > 10^6$ yr; thus, for any accretion rate $> 10^{-10} \smpy$, the thermal conditions in the bulk of the envelope are fixed at the start of mass transfer.  Calculations with constant mass removal rates of $10^{-5} $ and $10^{-9} \smpy$ were also run as a check, with negligible differences in the results.

\begin{figure}
	\plotone{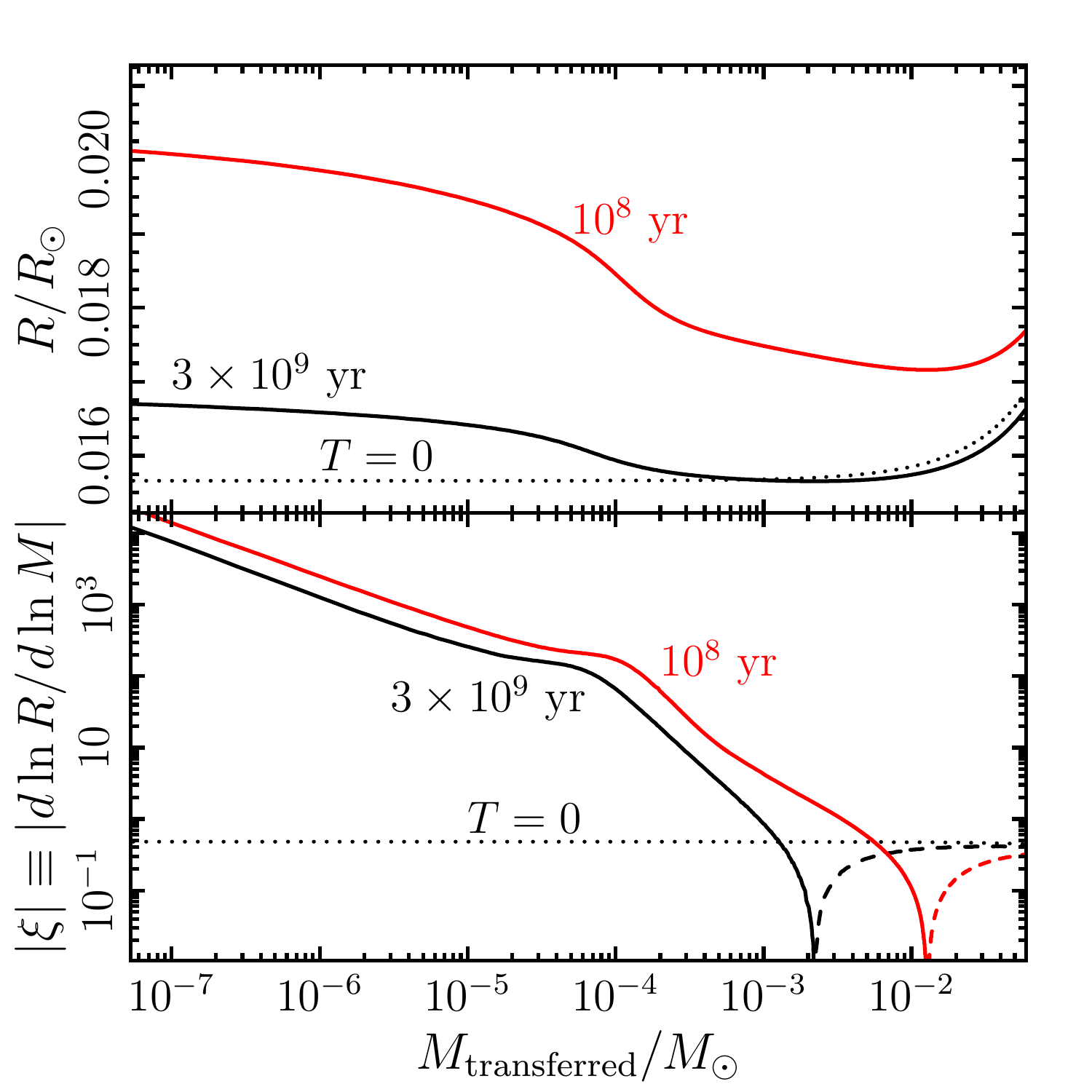}
	\caption{Radius (\emph{top panel}) and absolute value of the differential radial response (\emph{bottom panel}) vs.\ transferred mass for our two models.  Solid (dashed) lines in the bottom panel show positive (negative) values of $\xi$.  Dotted lines in both panels show zero-temperature analytic relations from \cite{naue72}.}
	\label{fig:randnvsm}
\end{figure}

The donor radius vs.\ removed mass is shown in the top panel of Figure \ref{fig:randnvsm} for our two models as labeled.  The younger model remains hotter and thus has a larger radius.  Both models have initial radii that are significantly larger than the zero-temperature analytic relation (\emph{dotted line}; \citealt{naue72}).  Once the $10^{-3}-10^{-2} \msol$ ideal gas, radiative layer has been stripped away, the degenerate He core expands upon further mass loss.

The bottom panel of Figure \ref{fig:randnvsm} shows the absolute value of the differential radial response, $\xi \equiv d \ln R / d \ln M$.  It is large and positive during the removal of the radiative H layer and approaches $0$ as this layer is exhausted.  As the radius of the mass-losing donor expands, $\xi$ becomes negative and eventually approaches the value for a zero-temperature low mass WD ($\xi \simeq -1/2$), shown as the dotted line.

In addition to depending on the donor radius and differential response, the actual accretion rate must also take into account the efficiency of angular momentum exchange between the accretor and the orbit \citep{nele01b,mns04}.  Assuming conservative mass transfer and angular momentum loss only due to gravitational wave radiation, the mass transfer rate is
\begin{eqnarray}
	\frac{\dot{M}_2}{M_2} = -\frac{32G^3}{5c^5} \frac{M_1 M_2 (M_1+M_2)}{a^4} \lp \frac{5}{6} + \frac{\xi}{2} - q - f(q) \rp^{-1} .
	\label{eqn:mdot}
\end{eqnarray}
The extra term $f(q)$, which accounts for the efficiency of angular momentum exchange \citep{vr88}, has little effect on the pre-merger evolution.  For our binary, the derived mass transfer rates for perfectly efficient and inefficient angular momentum feedback differ by less than $10\%$ while $\xi > 10$.  For $\xi=1$, their ratio is $1.82$.  Since mass transfer in our fiducial binary proceeds at all times via direct impact accretion instead of disk accretion \citep{ls75}, we assume that angular momentum feedback is inefficient for the remainder of this Letter.

% -----------------------------------------------------------
% -----------------------------------------------------------

\subsection{Accretion onto the C/O WD}

\begin{figure}
	\plotone{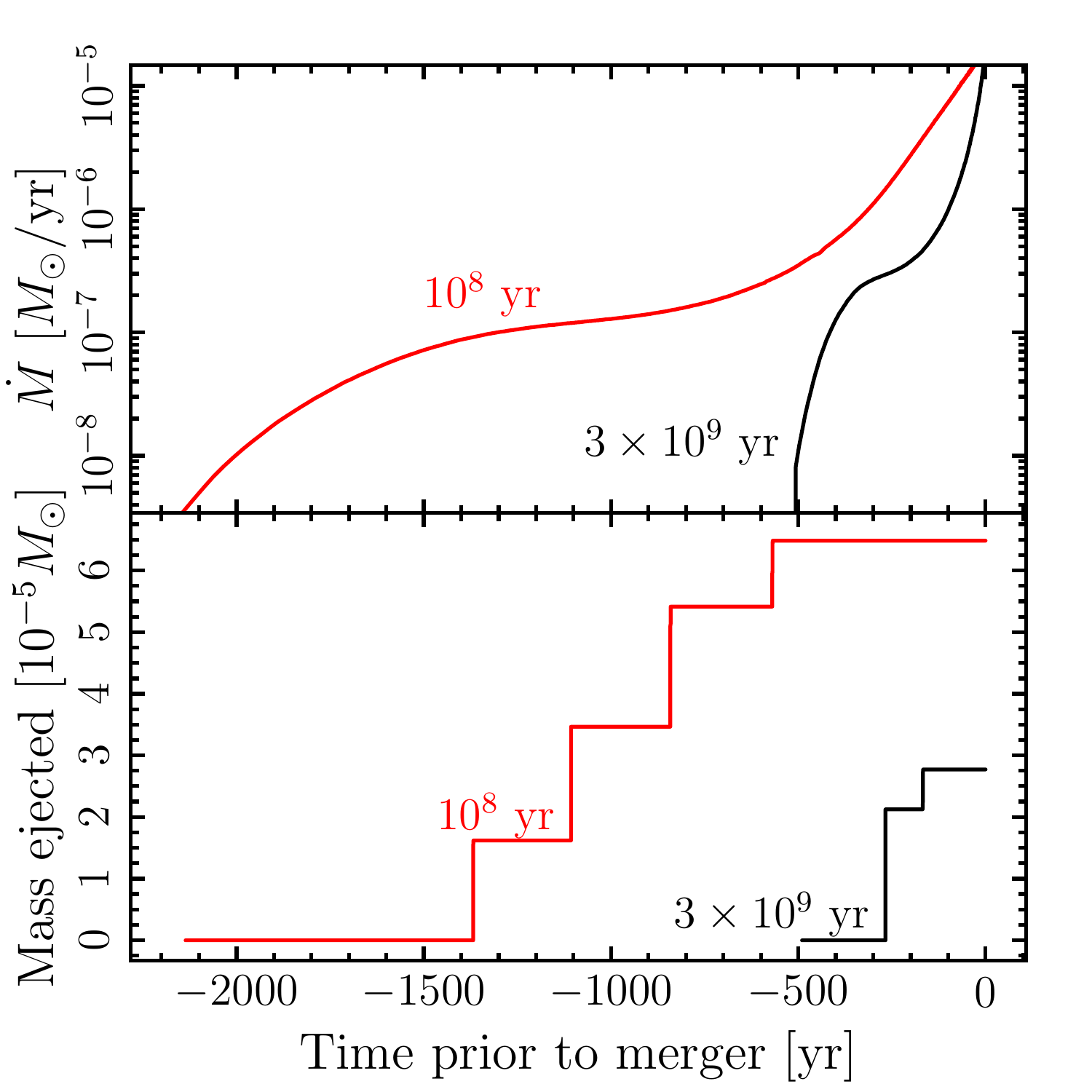}
	\caption{Derived mass transfer rate (\emph{top panel}) and mass ejected from the system (\emph{bottom panel}) vs.\ time prior to the merger for our two models.  Each system experiences multiple ejection episodes, yielding a total ejected mass of $3-6 \E{-5} \msol$.}
	\label{fig:mdotandmejvst}
\end{figure}

Using equation (\ref{eqn:mdot}), the donor's radial response shown in Figure \ref{fig:randnvsm}, and a relation between the donor's Roche lobe and the orbital separation \citep{eggl83}, we can derive the actual mass transfer rate as a function of time.  We show this accretion rate history in the top panel of Figure \ref{fig:mdotandmejvst} for our two models.  The accretion rate is plotted vs.\ the time prior to the WD merger, which we assume occurs when the accretion rate reaches the $1.0 \msol$ C/O WD's Eddington rate of $2\E{-5} \smpy$.\footnote{We assume that super-Eddington accretion will be driven from the system via dynamical friction with the donor.  The resulting evolution of the Roche overfill factor (e.g., \citealt{mns04}) implies a time of $1-10$ yr between the onset of super-Eddington accretion and the donor's disruption.  However, uncertainties such as the efficiency of angular momentum feedback and the actual Eddington rate may change this merger timescale.}

We model this time-dependent mass transfer onto a $1.0 \msol$ C/O WD in MESA, taking into account the changing composition of accreted material shown in Figure \ref{fig:0_4profiles}, but ignoring the C/O WD's own relatively small H- and He-rich layers for simplicity.  For faster convergence, the optical depth of the outer zone is moved inwards to $10^3$.  As H-rich material piles up on the surface of the C/O WD, the density and temperature at the base of the accreted layer increase until convective H-burning is ignited, as in a classical nova.  This causes the WD's radius to expand until it overflows its Roche radius.

We assume this material is driven out of the system by its interaction with the companion He WD via a common envelope (e.g., \citealt{livi90}), and that the ejection velocity is roughly equal to the He WD's circular velocity of $\simeq 1500$ km s$^{-1}$.  The bottom panel of Figure \ref{fig:mdotandmejvst} shows the mass ejection history prior to the disruption of the He WD.  A total of $3-6\E{-5} \msol$ of material is ejected from the binary in multiple ejections over the course of $200-1400$ yr prior to the merger.  The evolution of this ejecta and the swept-up ISM is the subject of the next section.

% -----------------------------------------------------------
% -----------------------------------------------------------

\section{Ejecta - ISM interaction}
\label{sec:ISM}

The interaction of an expanding shell of material with the surrounding ISM has been studied in detail with respect to SN remnants (e.g. \citealt{chev77}), classical and recurrent novae (e.g., \citealt{mb12}), and tidal tails from double WD mergers \citep{rk13a}.  Analytic solutions for such evolution are well-known; however, our situation is complicated by the existence of multiple ejection episodes, which will interact with previously shocked ISM.  We thus defer discussion of analytic results to future work and instead present numerical hydrodynamic simulations.

We employ a 1D, spherically symmetric, Eulerian hydrodynamics code that follows the zone-centered evolution of mass density, $\rho$, momentum density, $\rho v$, and energy density $\rho v^2/2 + \rho u$, where $u$ is the specific internal energy.  The equation of state only allows for ideal gas pressure because the medium is optically thin.  The mean molecular weight is assumed to be the solar value, $\mu = 0.6$, for the entire domain, which has a $10^{15}$ cm spatial resolution.  Fluxes are calculated with piecewise upwinded finite differencing.  Artificial viscosity is included via the prescription of \cite{tw79}.  An optically-thin cooling function, $\Lambda$, is included for $T > 10^4$ K by fitting to the results of \cite{gs07}.  The qualitative results in this section have been verified by a similar Lagrangian hydrodynamics code.

We calculate the interaction of our two fiducial binaries' ejecta with $10^2$ K ISM at three mass densities: $0.1$, $1$, and $10 \ m_p$ cm$^{-3}$, where $m_p$ is the proton mass.  Ejection episodes are initiated by increasing the density in the inner $r_{\rm pert} = 3\E{16}$ cm to a constant value of $3 M_{\rm ej} / 4 \pi r_{\rm pert}^3$, where $M_{\rm ej}$ is the ejecta mass, and setting the velocity to $v_{\rm ej} = 1500$ km s$^{-1}$, which is roughly the He WD's circular velocity.  The perturbed material's initial radius is chosen such that the ejecta is still essentially freely expanding.  Each ejection event is evolved until the next ejection, as prescribed by the ejection history in Figure \ref{fig:mdotandmejvst}, after which the inner zones are again perturbed in a similar manner.

\begin{figure}
	\plotone{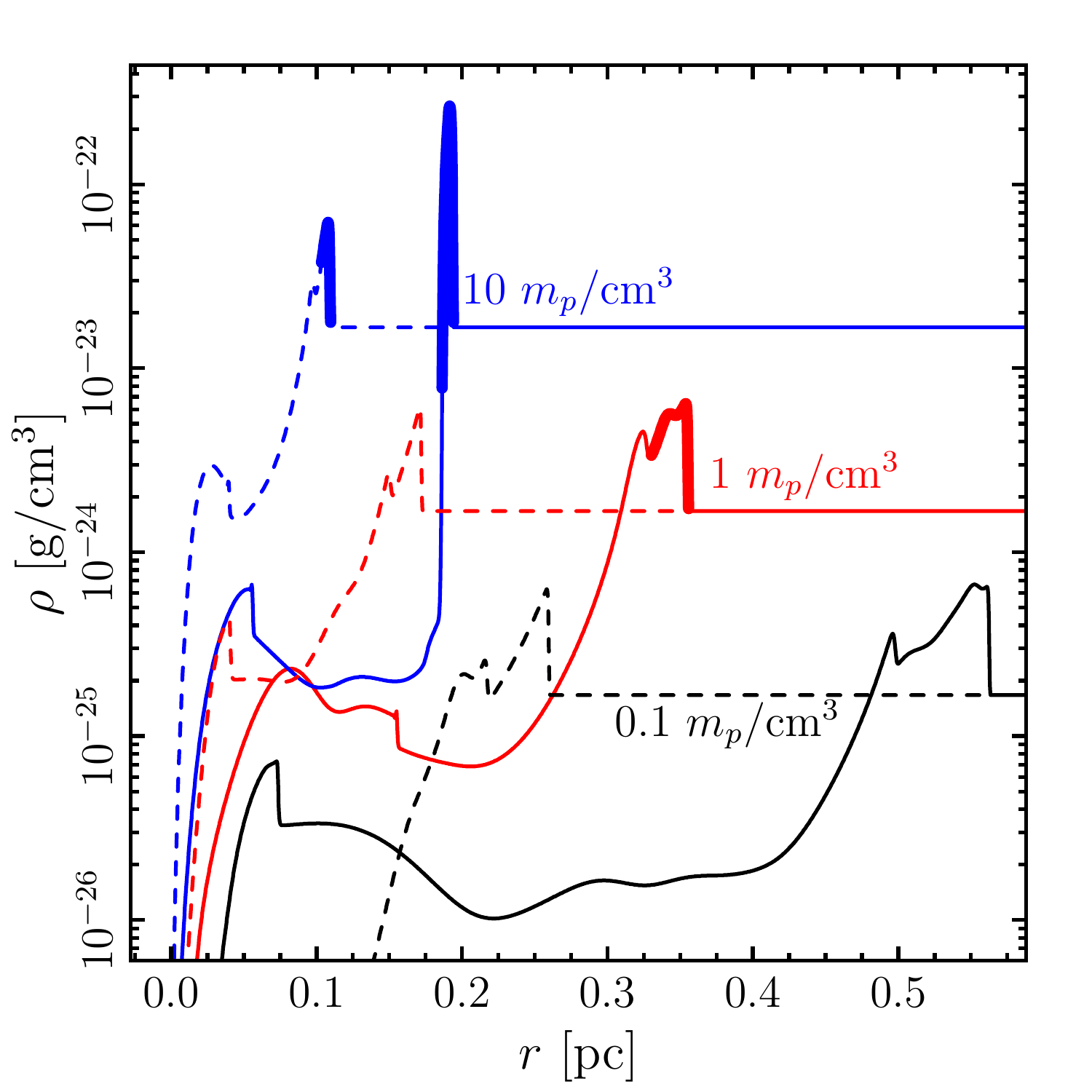}
	\caption{Mass density vs.\ distance at the time of explosion for our six models.  Solid lines correspond to the younger ($10^8$ yr) models; dashed lines show the older ($3\E{9}$ yr) models.  Labels denote the ambient ISM density.  Thick lines demarcate regions where neutral Na might exist in the CSM, as described in the text.}
	\label{fig:rhovsr}
\end{figure}

In Figure \ref{fig:rhovsr}, we show the mass density vs.\ distance from the SN at the time of merger for 6 simulations.  We find that when the SN Ia occurs, the outgoing shock is at a distance of $0.1-0.6$ pc and the just-shocked ISM has velocities of $50-300$ km s$^{-1}$.  These velocities correspond to post-shock temperatures of $6\E{4}-2\E{6}$ K.  Thus, without substantial cooling, there will be a negligible amount of blueshifted, neutral Na.  Significant cooling occurs in only one of our six simulations, which has the highest ISM density and the longest delay between the first ejection and merger.

However, there will likely be significant clumping in the ejecta (e.g., due to Rayleigh-Taylor instabilities), which is unresolvable in our 1D simulation.  Properly capturing the clumping will require multi-dimensional simulations, which we defer to future work.  Since the cooling timescale, $t_{\rm cool} \sim kT / n \Lambda$, is inversely proportional to density, clumping will enhance cooling in the shocked material and may allow for increased formation of neutral Na.  For now, we approximate regions where neutral Na may exist in the CSM by the conditions $T < 10^4$ K or $t_{\rm cool} < $ the simulation age at merger; these regions are shown in Figure \ref{fig:rhovsr} as thick lines.  Note that while clumping will alter the CSM's density, it will not significantly affect its position or velocity.

\begin{figure}
	\plotone{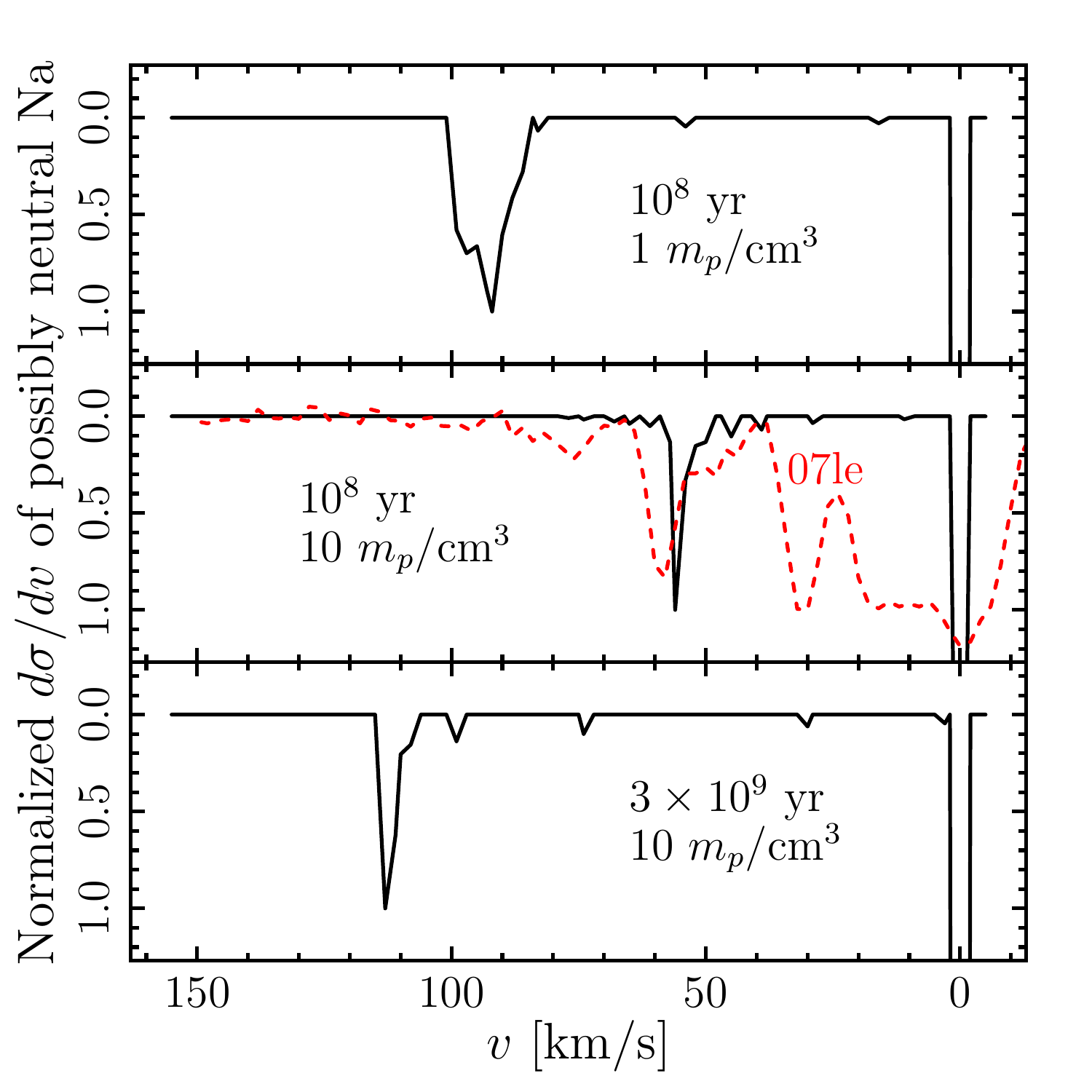}
	\caption{Differential column density per unit velocity, $d \sigma / d v$, vs.\ velocity of Na that might be neutral, as approximated in Figure \ref{fig:rhovsr}.  The normalization is such that the largest trough with non-zero velocity has a value of 1.  Thermal Doppler broadening is not included.  The velocity axis is reversed to match observational convention.  Values for the age of the He WD and the ambient ISM density are as labeled.  The red dashed line in the middle panel shows an average of SN 2007le's Na D absorption profiles 84 d after $B$-band maximum.}
	\label{fig:colvsv}
\end{figure}

In Figure \ref{fig:colvsv}, we show the normalized differential column density per velocity bin for regions that might contain neutral Na, using the same approximation as in Figure \ref{fig:rhovsr}.  The three models that contain any blueshifted, possibly neutral, Na are shown in the three panels as labeled.  This metric approximates the neutral Na absorption line profile and shows systemic velocity offsets of $50-120$ km s$^{-1}$.  For comparison, the red dashed line in the middle panel shows the normalized average of the Na D$_1$ and D$_2$ absorption profiles for SN 2007le 84 d after $B$-band maximum \citep{simo09}.

The velocity profiles of the possibly neutral Na resemble those seen in observations, and the Na column densities ($4\E{11}-4\E{12}$ cm$^{-2}$) overlap with those derived from SNe Ia showing variable absorption lines.  However, our column densities are a factor of a few higher than the observed mean of \cite{ster11}.  Given our ad hoc inclusion of the effects of clumping and the unconsidered complications of time-dependent photoionization and recombination following the SN, our derived neutral Na column densities are merely suggestive.  Future multi-dimensional work will enable more accurate predictions and possibly correct this mismatch.

No narrow Na absorption has yet been detected in a SN Ia in an elliptical galaxy without obvious dust lanes \citep{fole12a}.  Our results agree with this finding for several reasons.  Elliptical galaxies have lower ISM densities, which yield a smaller column of shocked material at the time of the explosion.  Furthermore, the lower ISM density implies both lower post-shock densities as well as less shock deceleration, which means higher CSM temperatures when the SN Ia occurs.  These factors increase the cooling timescale and decrease the amount of neutral Na.

% -----------------------------------------------------------
% -----------------------------------------------------------

\section{Conclusions}
\label{sec:conc}

In this Letter, we have considered the effect of the H-rich layer that surrounds a He WD on the He WD's interaction with a C/O WD companion prior to a SN Ia.  We have calculated its structure (Section 2), its impact on mass transfer and its ejection (Section 3), and the ejecta's interaction with the surrounding ISM (Section 4).  We have found that if a SN Ia occurs when the He and C/O WDs merge, the characteristics of the CSM at the time of the explosion match recent observations of neutral Na surrounding $10-30\%$ of SNe Ia in spiral galaxies.  We have also found that the lower ISM density in elliptical galaxies inhibits the formation of significant neutral Na in the CSM, which may be why these features have not been detected in such SNe Ia.

SN ejecta have been observed to collide with surrounding CSM in several SNe Ia (e.g., \citealt{hamu03,dild12,silv13b}).  This interaction requires significant material within $\lesssim 10^{16}$ cm, which is difficult to produce in our model unless the SN Ia occurs $<2$ yr after an ejection event.  More detailed modeling and inclusion of the super-Eddington accretion phase just before the merger may help to shed light on these observations.

While our results are promising, future studies are necessary to strengthen the findings.  Further work will include an exploration of a range of WD masses and ages, simulations of the ejecta - ISM interaction in multiple dimensions, which will allow for clumping and non-spherical ejection, and calculations of the time-dependent photoionization and recombination after the SN Ia's UV flash.  The effects of tidal heating, which will likely be significant for these extremely close binaries (e.g., \citealt{piro11a}), should also be considered.

% -----------------------------------------------------------
% -----------------------------------------------------------

\acknowledgments

We thank Jason Dexter, Dan Kasen, Rodolfo P\'{e}rez, Eliot Quataert, Cody Raskin, and Jeff Silverman for discussions.  KJS is supported by NASA through Einstein Postdoctoral Fellowship grant number PF1-120088 awarded by the Chandra X-ray Center, which is operated by the Smithsonian Astrophysical Observatory for NASA under contract NAS8-03060.

% -----------------------------------------------------------
% -----------------------------------------------------------

% -----------------------------------------------------------
% -----------------------------------------------------------

\end{document}